\documentclass{eas}
\usepackage{graphicx}
\def\gtsim {>\kern-1.2em\lower1.1ex\hbox{$\sim$}~}   
\def\ltsim {<\kern-1.2em\lower1.1ex\hbox{$\sim$}~}   
\def \apj {ApJ}
\def \apjs {ApJS}

\def \mnras {MNRAS}

\TitreGlobal{Proceedings of the CRAL-Conference Series I "Chemodynamics: from first stars to local galaxies"}
\Editors{eds. Emsellem, Wozniak, Massacrier, Gonzalez, Devriendt, Champavert}
\begin{document}

\title{Simulations of Cosmic Chemical Enrichment with Hypernova} 
\author{Chiaki Kobayashi}\address{National Astronomical Observatory of Japan, 2-21-1 Osawa, Mitaka-shi, Tokyo 181-8588, Japan; chiaki@th.nao.ac.jp}
\vspace*{-2mm}
\begin{abstract}
We simulate cosmic chemical enrichment with a hydrodynamical model including supernova and hypernova feedback.
We find that the majority of stars in present-day massive
  galaxies formed in much smaller galaxies at high redshifts, despite their late assembly times. The
  hypernova feedback drives galactic outflows efficiently in low mass galaxies, and these
  winds eject heavy elements into the intergalactic medium.  The ejected
  baryon fraction is larger for less massive galaxies, correlates well with
  stellar metallicity. 
 The observed mass-metallicity relation is well reproduced 
as a result of the mass-dependent galactic winds.
We also predict the cosmic supernova and gamma-ray burst rate histories.
\end{abstract}
\maketitle
\vspace*{-2mm}
\section{Introduction}
\vspace*{-2mm}

While the evolution of the dark matter
is reasonably well understood, the evolution of the baryonic component is much
less certain because of the complexity of the relevant physical processes, such
as star formation and feedback.
With the commonly employed,
schematic star formation criteria alone, the predicted star formation rates (SFRs)
are higher than what is compatible with the observed luminosity
density.  Thus feedback mechanisms are in general invoked to reheat gas and
suppress star formation. 
We include both supernova and hypernova feedback in
our hydrodynamical model in this paper (see Kobayashi, Springel \& White 2006b, hereafter K06b,  for the details).

\vspace*{-2mm}
\section{Hypernovae}
\vspace*{-2mm}

Supernovae eject not only thermal energy but also heavy elements into interstellar medium.
Thus the star formation history is imprinted in the chemical abundances of stars and galaxies.
Different types of supernovae, i.e., Type II and Ia supernovae (SNe II and Ia), produce different heavy elements with different timescales.
Recently, it is found that hypernovae (HNe), which have more than ten times larger explosion energy ($E_{51}\gtsim10$), produce a certain amount of iron.

Kobayashi et al. (2006a) calculated the nucleosynthesis yields for wide ranges of metallicity and energy, based on the light curve and spectra fitting of individual supernovae.
With these yields, the elemental abundance ratios are in good agreement with observations from oxygen to zinc.
Especially, the observed abundance of Zn ([Zn/Fe] $\sim0$) can be explained only by a large contribution of HNe.
We adopt these yields and energies in our hydrodynamical model.

\begin{figure}[t]
\begin{center}
\includegraphics[width=5.5cm]{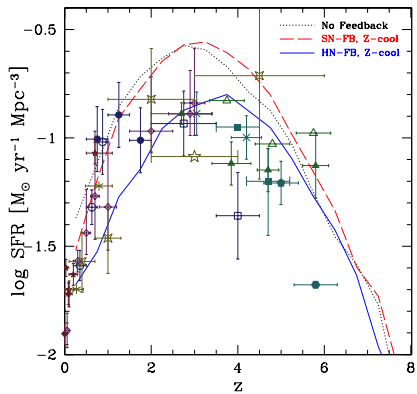}
\caption{\label{fig:sfr}
Cosmic star formation rates for no feedback case (dotted line) and for the SN (dashed line) and HN (solid line) feedback with the metal-dependent cooling.
See K06b for the references of Rest-frame U-band, IR, and H$\alpha$ observations.
Rest-frame UV observations are plotted with dust correction (Steidel et al. 1999).
}
\end{center}
\end{figure}

\vspace*{-2mm}
\section{Cosmological Simulations}
\vspace*{-2mm}

We simulate the evolution of gas and stellar systems and the chemical enrichment from SNe II, SNe Ia, and HNe, from the cosmological initial condition with $H_0=70$ km s$^{-1}$ Mpc$^{-1}$, $\Omega_m=0.3$, $\Omega_\Lambda=0.7$, $\Omega_{\rm b}=0.04$, $n=1$, and $\sigma_8=0.9$.
We use an SPH code GADGET-2 by Springel (2005), and introduce the metal-dependent cooling rates (Sutherland \& Dopita 1993) and chemical enrichment scheme by Kobayashi (2004).
The initial condition is calculated in a $10 h^{-1}$ Mpc cubic box with periodic boundary conditions with $N_{\rm DM}=N_{\rm gas}=96^3$.
We adopt the Salpeter IMF, the star formation timescale $c_*=0.1$, and the feedback neighbors $N_{\rm FB}=405$.

Figure \ref{fig:sfr} shows the cosmic SFRs that are directly measured from the
ages of stellar particles for the low resolution simulations.
Supernova feedback (dashed line) decreases the SFR from
$z\sim3$, but metal-dependent cooling increases as large as no feedback case. 
For comparison, Figure \ref{fig:sfr} also shows observational estimates of the
cosmic SFR density at different epochs. Note that these determinations are
derived from the observed luminosity densities, and this
involves uncertainties from dust extinction and completeness, as well as from
the IMF.
If we include hypernova feedback (solid line), the SFR starts to be
suppressed from $z\sim6$ onwards, and is overall smaller by a factor of $3$ at
$0\ltsim z \ltsim3$.  The resulting SFR is in broad agreement with the
observations that show a peak of $\log({\rm SFR}/[{\rm M}_\odot{\rm yr}^{-1}])
\sim -1$ at $z\sim4$. 
Since the metallicity dependence of the cooling rate clearly appears for [Fe/H]
$\gtsim -2$, metal-dependent cooling does not play such a prominent role in
the HN feedback case, and enriched gas can remain hot without forming stars.

Without feedback,
$25\%$ of baryons turn into stars, which is too large compared with
observational estimates.  With SN feedback, the stellar
fraction reduces to $10-15\%$, which may be consistent with observation.
Recently, the observational estimate has been reduced to less than $10\%$
(Fukugita \& Peebles 2004), which may require larger feedback.  The larger energy ejection
by HNe could provide a solution.  The present mean gas metallicity is [Fe/H]
$\sim -1.3$ for SN feedback, and is reduced to $\sim -1.8$ for HN feedback.
The mean stellar metallicity is almost solar for the SN feedback, and becomes
sub-solar for the HN feedback.

\begin{figure}[t]
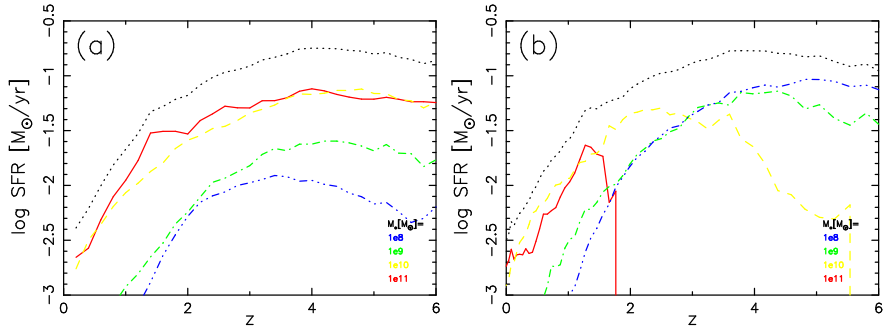

\begin{center}
\includegraphics[width=4.3cm,angle=-90]{kobayashi_fig2a.ps}
\includegraphics[width=4.3cm,angle=-90]{kobayashi_fig2b.ps}
\caption{\label{fig:sfr_deco}
The cosmic SFR for the different galaxy mass with the stellar mass of $10^{11}$ (solid line), $10^{10}$ (dashed line), $10^9$ (dot-dashed line), and $10^8 M_\odot$ (dot-dot-dot-dashed line). The dotted line shows the total.
Galaxies are identified by FOF at $z=0$ (a) and at each redshift (b), respectively.
}
\end{center}
\end{figure}

When and where do stars form? To answer this question,
we break up the cosmic SFR history according to galaxy mass.
In Figure \ref{fig:sfr_deco}a, the galaxies have been identified by FOF at $z=0$ for this
plot, and thus these SFRs correspond to the age distribution of stars in the
galaxies.  For all galaxy masses, the SFRs show a peak around $z\sim3-4$, and
the majority of stars are as old as $\sim 10$ Gyr.  On the other hand, in
Figure \ref{fig:sfr_deco}b, we identify galaxies at each redshift, 
which are comparable to the observations of high redshift galaxies.
This shows that most stars have formed in low-mass galaxies
with $10^{8-9}{\rm M}_\odot$ at high redshift $z \gtsim 3$. 
$10^{10}M_\odot$ galaxies exist at high redshift $z \ltsim 5$, but $10^{11}M_\odot$ galaxies appear only after $z \sim 2$.
From these two
figures we conclude that most stars have formed in dwarf galaxies 
before they merge to massive galaxies in our simulation.
As a result of the
hierarchical clustering of dark matter halos, such old stars belong to massive galaxies at low redshifts.

\begin{figure}[t]
\begin{center}
\includegraphics[width=11cm]{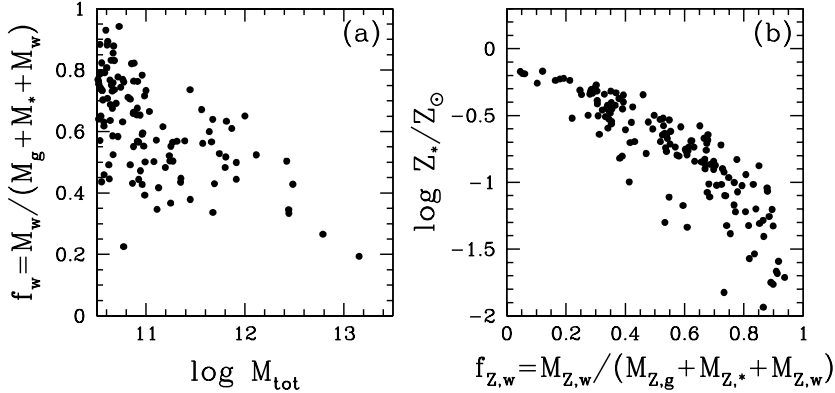}
\caption{\label{fig:wind}
The wind fraction against the total mass (a), and the stellar metallicity against the ejected metal fraction (b).   
}
\end{center}
\end{figure}

How are heavy elements ejected from galaxies to the IGM?  In the simulation,
we can trace the orbit of gas particles over time. Exploiting this, we define
as wind particles those that are not in galaxies now, but have been in
galaxies before.  
In this simulation, $\sim 10\%$ of baryons turn into stars, $\sim 10\%$
of the gas stays in galaxies ($\sim 8\%$ is hot), and $\sim 20\%$ is ejected
as galactic winds.  The rest, half of the baryons, never accretes onto galaxies.

When we follow the orbits of gas particles, we can also examine from which
galaxies the wind gas particles are ejected. This allows a measurement of
the ejected wind mass from each galaxy.  
In Figure \ref{fig:wind}a, a clear relation is found between the wind fraction 
and the total mass. 
Winds are efficiently ejected from small
galaxies, with $\sim 80\%$ of accreted baryons being ejected from $M_{\rm
  tot}\sim 10^{11}{\rm M}_\odot$ galaxies.
A similar relation is also found for the ejected metal fraction, i.e.~the ratio between the wind metal mass to the total metal mass.
It is interesting that the wind
fraction and the ejected metal fraction correlate well with the stellar metallicity (Fig.\ref{fig:wind}b).  Based on
this finding, we conclude that the origin of the mass-metallicity relation can
be explained with galactic winds.

\begin{figure}[t]
\begin{center}
\includegraphics[width=5.5cm]{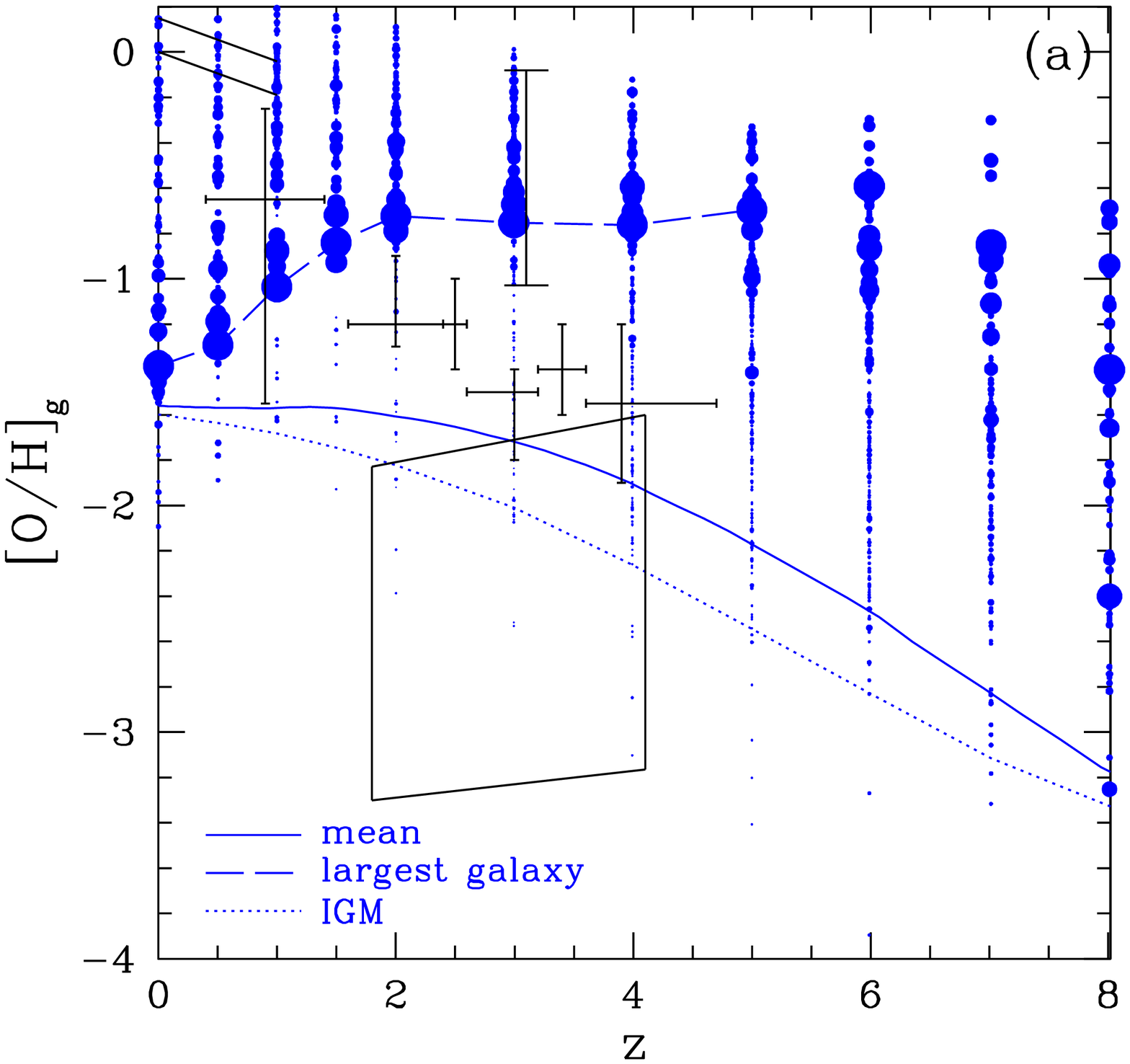}
\includegraphics[width=5.5cm]{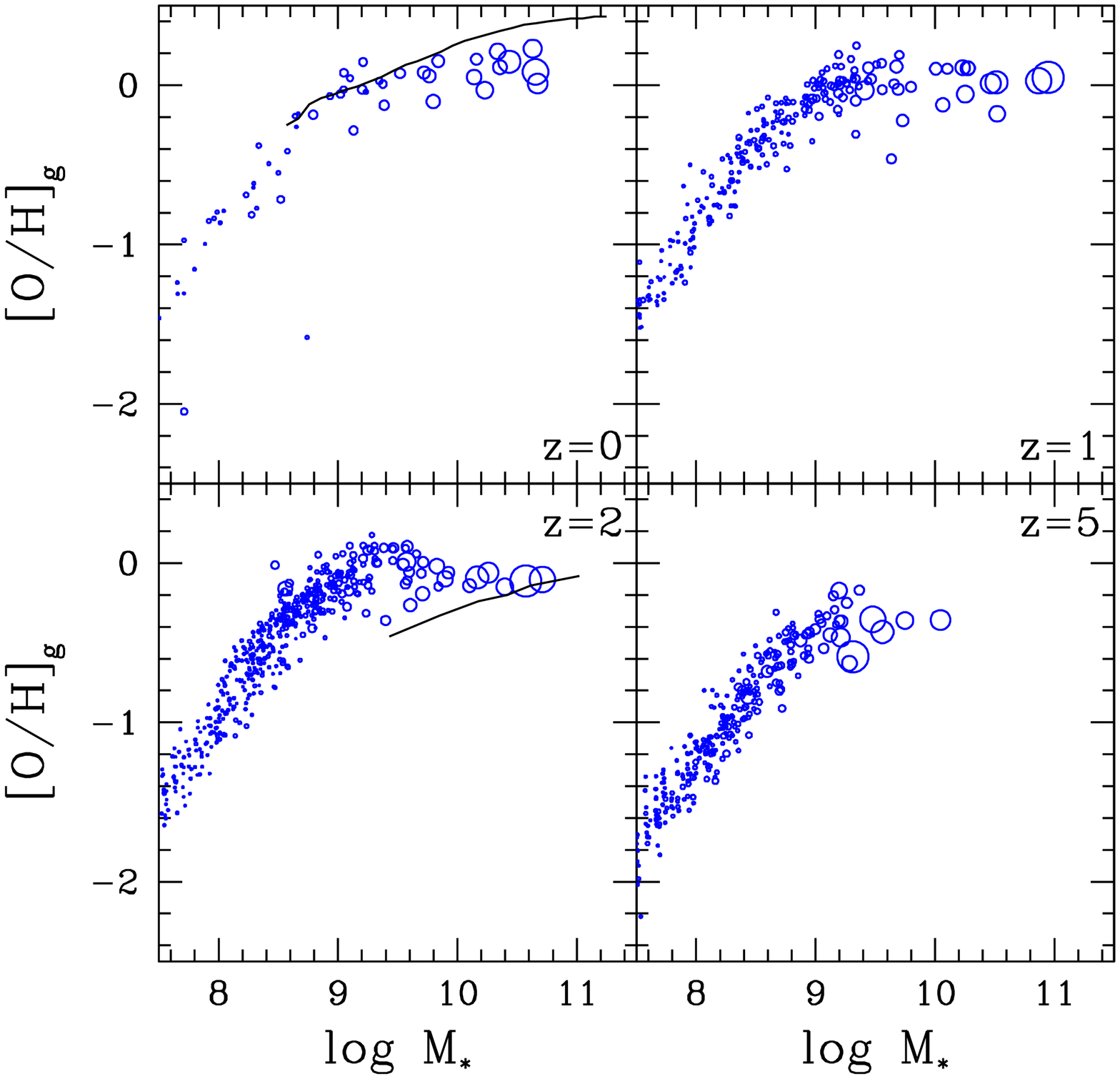}
\caption{\label{fig:feh}
(a)
Redshift evolution of oxygen abundances of gas.
The points show individual galaxies with the size representing the size of galaxies.
The solid, dashed, and dotted lines are for the mean, the largest galaxy, and the IGM, respectively.
See the text for the observational data (errorbars).
(b)
  Mean metallicities of cold gas ($T<10^4$ K) within 10 kpc,
  plotted against the total stellar mass at $z=0,1,2,$ and $5$.
See the text for the observational data (solid lines).
}
\end{center}
\end{figure}

The metal enrichment timescale depends on the environment.
Figure \ref{fig:feh}a shows the evolution of oxygen abundances in the gas
phase. 
In large galaxies,
  enrichment takes place so quickly that [O/H] reaches $\sim -1$ at $z\sim7$,
  which is consistent with the sub-solar metallicities of the Lyman break
  galaxies (large errorbar at $z=3$, Pettini et al. 2001).  The low metallicities of DLA systems (errorbars, Prochaska et al. 2003) are also consistent with our
  galaxies, provided these systems are dwarf galaxies or the outskirts of
  massive galaxies.  The low [C/H] of the IGM (box, Schaye et al. 2003) can be explained if the IGM is
  enriched only by SNe II and HNe.  The average metallicity of the universe
  reaches [O/H] $\sim -2$ and [Fe/H] $\sim -2.5$ at $z\sim4$ 
, but
  reaches the same values at $z\sim3$ 
in the IGM.

In galaxies, metallicity of the cold gas increases with galaxy mass (Fig.\ref{fig:feh}b), 
  which is comparable to observations with a large scatter (solid line at $z=0$, Tremonti et al. 2004; $z=2$, Erb et al. 2006).
  The central cold gas shows a relation between galaxy mass
  and metallicity with shallower slope than observed in emission-line
  galaxies.  For the stellar population, the observed mass-metallicity
  relation is well reproduced (Kobayashi \& Arimoto 1999; Pahre et al. 1998; Gallazzi et al. 2005), and originates in mass-dependent galactic winds.
  These relations are present since $z \sim 5$.

\vspace*{-2mm}
\section{Cosmic Supernova Rates}
\vspace*{-2mm}

\begin{figure}[t]
\begin{center}
\includegraphics[width=8cm]{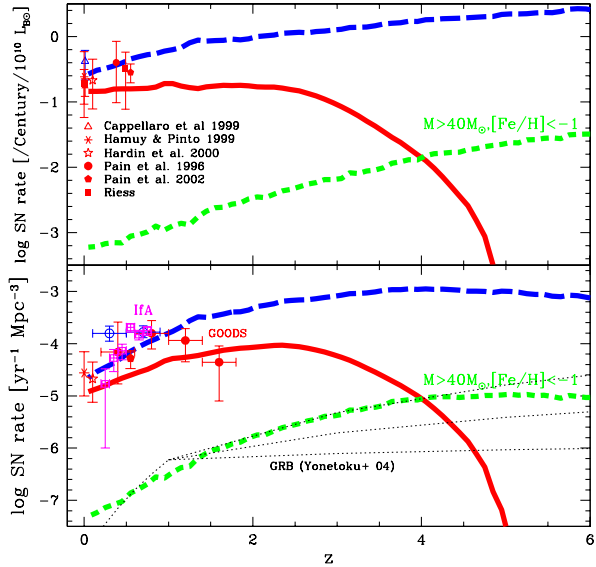}
\caption{\label{fig:snr}
The cosmic rates for SNe II (long-dashed line), SNe Ia (solid line), and GRB (short-dashed line) per luminosity (SNu, upper panel) and volume (lower-panel).
The dotted line shows the observation of GRB rates (Yonetoku et al. 2004).
The open triangle and open circle are for the observations of SNe II rates, and the other symbols are for SNe Ia.
See Kobayashi \& Nomoto (2006, in preparation) for the observational data sources.
}
\end{center}
\end{figure}

We provide the cosmic supernova rates that is calculated with the cosmological  chemodynamical simulations in Figure \ref{fig:snr}.
Recently, large scale and deep surveys can give the SNe Ia rates as a function of redshift (\cite{dah04}, \cite{bar05}).
From the evolution of SN Ia rates, the delay time of SNe Ia have been constrained as a few Gyr (\cite{man06}).
However, the theory of the SN Ia progenitor (i.e., a binary system of a white dwarf) is not simple and the evolution of the binary system should be taken into account.
The delay time corresponds to the lifetime of the binary companion, and the shortest time of SNe Ia is much shorter as $\sim 0.1$ Gyr for the double-degenerate system (\cite{tut94}), $\sim 0.3$ Gyr for \cite{mat01}'s model, and $\sim 0.5$ Gyr for our single-degenerate system (\cite{kob98}).

In addition, because the occurrence of SNe Ia depends on the metallicity in our SN Ia progenitor scenario, the SNe Ia rate depends on the environments, i.e., the star formation and chemical enrichment histories of host galaxies.
Therefore, not with one zone model of the Universe but with the chemodynamical simulations, we should predict the cosmic supernova rates.
In our simulations, the effects of hypernovae are included, and we can also provide the cosmic gamma-ray burst (GRB) rates assuming that the progenitors of GRB are massive ($\ge 40 M_\odot$) and low-metal ([Fe/H] $\le-1$) hypernovae.
Our predictions are in good agreement with observations.

\vspace*{-2mm}


\end{document}